\documentstyle[prl,floats,preprint,aps,epsfig]{revtex}

\def\psihat{{\hat \psi}}
\def\(({\left(}
\def\)){\right)}
\def\be{\begin{equation}}
\def\ee{\end{equation}}

 \tightenlines 

\begin{document}

\title{Stretching an heteropolymer}

\author{D.Bensimon$^{1,2}$, D.Dohmi$^{3}$ and M.M\'ezard$^4$}


\address{(1) Laboratoire de Physique Statistique de l'ENS \footnote{
Laboratoire associ\'e au CNRS et
aux universit\'es Paris VI et VII}\\
24 rue Lhomond, 75231 Paris Cedex 05, France\\
(2) Dept. of Complex Systems, Weizmann Institute, Rehovoth, Israel\\
(3) Dept. of Physics, Univ. Rabat, Rabat, Morocco\\
(4) Laboratoire de Physique Théorique de l'ENS \footnote{Laboratoire propre du CNRS,
associ\'e \`a l'ENS et \`a l'Universit\'e de Paris XI}\\
24 rue Lhomond, 75231 Paris Cedex 05, France}

\maketitle
\begin{abstract}
We study the influence of  some quenched disorder in the sequence
of monomers on the entropic elasticity of long polymeric chains. Starting from the
 Kratky-Porod model, we show numerically
that some randomness in the favoured angles between 
successive segments induces a change in the  elongation versus
force characteristics, and  this change can be well described by a
simple renormalisation of the elastic constant. The effective coupling
constant is computed by an analytic study of the
low force regime. LPTENS 97/51.

\end{abstract}


New tools have recently been developped for the manipulation of single
molecules, in particular DNA\cite{finzi,cluzel,strick,block}, protein fibers,
such as titin\cite{busta,simmons,gaub1} and polymers\cite{damm95}.
In these experiments, the fibers are stretched by various means
(optical or magnetic tweezers, flexible microscopic cantilevers,
stokes drag) and their extension is measured.  The various models
(e.g., the Freely Jointed Chain (FJC) or Worm-like Chain (WLC) models) used to
analyse the data are based on the elastic theory of an homogenous
polymer stretched in its entropic regime.  Because the actual
molecules, particularly proteins and DNA, are in fact heteropolymers, it is
interesting to study how  their intrinsic quenched
randomness modify their elastic behaviour\cite{gardner}. This could be particularly
relevant to possible stretching experiments on single strand DNA or
proteins in denatured conditions.

In this paper we wish to understand the qualitative new features introduced by
randomness. We shall not consider a particularly realistic type of disorder
(this should be tailored for each type of molecule). Rather
we want to start from the usual 
Kratky-Porod model  \cite{cantor} of an elastic chain, and compute
the effects of the introduction of a certain simple type of disorder, namely some
 preferred orientations between successive links. 
 In the following we compute analytically the effective (disorder averaged)
persistence length at low extensions and use numerical methods
(transfer matrix and Monte-Carlo) to solve for the behaviour of the
chain at all extensions. Surprisingly, it appears that the full
elastic response of this random chain is similar to the response of an
homogenous chain with a different elastic constant. The effective elastic constant
can be determined from the calculated low extension effective
persistence length.

In the Kratky-Porod model, the energy ${\cal E}$
of an homogenous polymer chain, consisting of $N$ segments of size $b$
(maximal extension $L = N b$) and orientation given by the unit vectors $\vec{t}_i$
is equal to:

\be   \label{e1}
\frac{{\cal E}}{k_B T} =
-K \sum_i \vec{t_i} \cdot \vec{t}_{i+1} - H \vec{t_{i}}. \hat z 
= -K \sum_i \cos \alpha_i - H \sum_i \cos \theta_i  \ ,
\ee

where $\alpha_i$ is the angle between the successive segments $i-1$ and
$i$ and $\theta_i$ is the angle between segment $i$ and the $z$-axis, the
direction of the stretching force $F$.
 One recognises the analogy between polymer chains and a
magnetic system: the ferromagnetic chain of vector
spins in a magnetic field \cite{fisher}. The magnetic field $H$ is related to the
applied force $F$ through $ H = F b / k_B T$, while the ferromagnetic coupling $K$ is
related to the stiffness or persistence length of the chain $\xi_T$ by
$K= \xi_T / b$. The FJC model (obtained by setting $K=0$) is equivalent to
the behaviour of an ensemble of non-interacting spins. The WLC
model is the continuum limit of Eq.(\ref{e1}): $ b \to 0$,
$\xi_T$ and $F$ finite (i.e.: $K \gg 1 \gg H$):

\begin{eqnarray}
\label{e2}
{{\cal E} \over k_B T} = 
{\xi_T \over 2} \int_0^L \left({ d \vec t \over ds}\right)^2 ~ ds
- {F \over k_B T} \int_0^L \vec t. \hat z \ ds  \ .
\end{eqnarray}

As the only non-dimensional parameter in Eq.(\ref{e2}) is $F \xi_T /
k_B T$, the relative extension $x = <\cos \theta > = l/L$ (equal to the magnetization
density in the spin language) will be a
function of that parameter only, or inverting:

\begin{eqnarray}
\label{e2a}
 F= (k_B T / \xi_T) ~ g(x)  \ .
\end{eqnarray}
The function $g(x)$ (valid for any homopolymer) can be computed from a Hamiltonian formulation
by solving for the ground state of a quantum dumbel \cite{fixman}. It is known to reproduce
with good accuracy the experimental data on DNA \cite{marko}.

In the following we shall consider a random version of the
Kratky-Porod Model just described, where the energy of the chain is:

\begin{eqnarray}  \label{e3}
{{\cal E} \over k_B T}  
= -K \sum_i \cos (\alpha_i - \psi_i) - H \sum_i \cos \theta_i \ .
\end{eqnarray}

Here $\psi_i$ is a preferred random orientation between the successive
segments $i-1$ and $i$ along the chain. The angles $\psi_i$ are {\it quenched},
independent and identically distributed, random variables, with probability distribution
$P(\psi_i)$. It is easy to obtain the spin-spin correlation function at
zero field \cite{fisher}:

\begin{eqnarray}   \label{e4}
<\vec t_{i} . \vec t_{i+l}> = <\vec t_{i}. \vec t_{i+l-1}>
 f(\psi_{i+l}) 
= \prod_{j=i+1}^{i+l} f(\psi_j) \ ,
\end{eqnarray}
where:
\begin{eqnarray}   \label{e5}
f(\psi) &=& { \int d\alpha  \ \sin \alpha \cos \alpha \ e^{K \cos (\alpha - \psi)}
\over \int  d\alpha \ \sin \alpha \  e^{K \cos (\alpha - \psi)} } \ .
\end{eqnarray}

The correlation (\ref{e4}) decays at large argument as $\exp(-r b/\xi_{eff})$,
where the effective correlation length is
${b / \xi_{eff}}=-\overline{\log(f)}$ (throughout the paper,
the overline denotes the
average over the distribution $P(\psi)$ of the random angle $\psi$).
This result for the effective correlation length can be simplified in
the continuum limit where $b \to 0$. In this limit the stiffness scales
as $K=\xi_T/b$. We shall make the reasonable assumption that the
distribution of disorder scales as $P(\psi)=\hat P(\psihat)/\sqrt{b}$ where
$\psihat=\psi/ \sqrt{b}$, and that the angles $\psi_i$ are all
positive. Using the fact that the angle $\alpha$ also scales
as $\sqrt b$ in (\ref{e5}), one gets to first order in $b$:
\be
f(\psi)=1-{b \over 2 \xi_T} G(y)
\label{fexpansion}
\ee
where $y=\sqrt{\xi_T} \psihat$ and the function G is:
\be 
G(y)=3 +y^2-\((1+\sqrt{{\pi \over 2}} \ y e^{y^2/2} (1+{\rm Erf}[y/\sqrt{2}])\))^{-1}
\ee
In terms of this function $G$ the effective persistence length is given by:
\be
{1 \over \xi_{eff}}={1 \over 2 \xi_T}  \int d \psihat\  \hat P(\psihat)
 G(\psihat \sqrt{ \xi_T})
\label{xieff}
\ee

In the case where the disorder is small, which means that the
typical values of the angles $\psi_i$ are much smaller than $\sqrt{b/\xi_T}$,
one can get a general result for the effective persistence length by expanding
the function $G$ at small arguments:
\be
{1 \over \xi_{eff}}={1 \over \xi_T} +\sqrt{{\pi \over 8}} \overline{\psihat}
+\((1-{\pi \over 4}\)) \overline{\psihat^2}+ ...
\label{weak}
\ee

In the reverse case of strong disorder one gets:
\be
{1 \over \xi_{eff}}={\overline{\psihat^2} \over 2}+{3 \over 2 \xi_T} +...
\label{strong}
\ee

In fig. 1 we plot the effective persistence length in
units of the persistence length of the pure system, $\xi_{eff}/\xi_T$, as
a function of the ratio $r=\xi_T/\xi_d$ ($r$ is a measure of the strength of
disorder), for a half-gaussian distribution of disorder
$
P(\psi) = 
\sqrt{2 \xi_d/ \pi b} \exp(- \xi_d \psi^2 / 2 b) \ \ (\psi>0)$. The
previous asymptotic results show that $\xi_{eff}/\xi_T$ behaves as
$1-\sqrt{r}/2$ at small $r$ and as $2/r$ at large $r$.
These results differ from the heuristic arguments often found in the 
litterature\cite{trifonov,schellman} which appear to consider the 
annealed disorder case with gaussian disorder, yielding:
\begin{equation}    \label{naive}
{1 \over \xi_{eff}^{an}}  \sim {1 \over \xi_d} + {1 \over  \xi_T }
\end{equation}
or $\xi_{eff}/\xi_T=1/(1+r)$, which is also shown in fig.1.

\begin{figure}
\centerline{
\hbox{\epsfig{figure=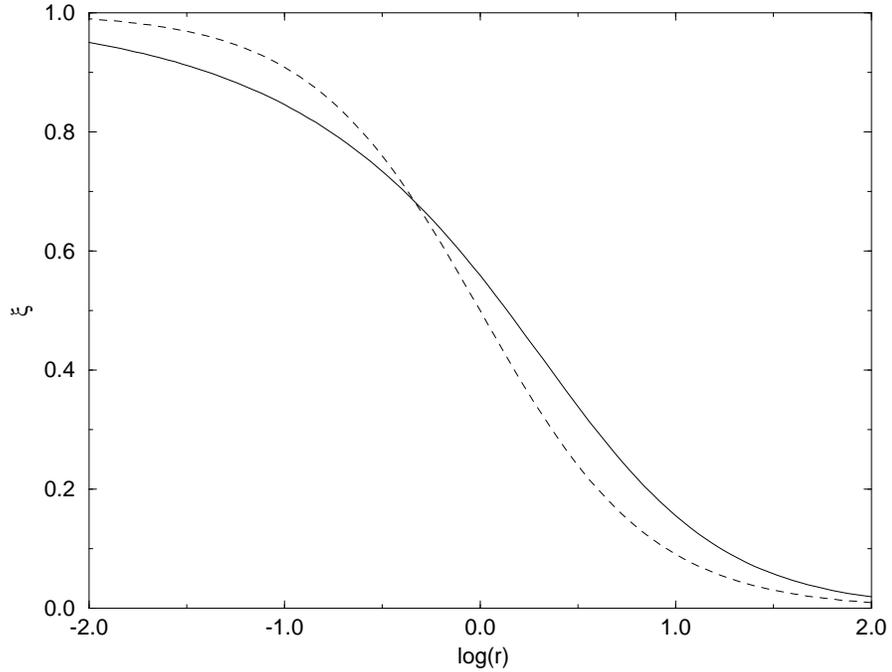,width=10cm,angle=-90}}
}
\caption{The ratio of the effective persistence length to the one
of the pure homopolymer, $\xi_{eff}/\xi_T$, versus the strength of
the disorder, measured by the ratio $r=\xi_T/\xi_d$,
 in logarithmic scale. The full curve is the
exact result from (\ref{xieff}) for a half gaussian disorder (see text),
the dashed line is the result of the simple heuristic formula (\ref{naive}).}
\label{fig1}
\end{figure}

Very often, the actual measurements do not have a direct access to the
correlation length, but they deduce it from low force measurements. In the limit
of small forces, the relative extension is linear in the force with 
a slope:
\be
S \ = \ {\partial x \over \partial F}(F=0) \ = \ {b \over N k_B T}\sum_{i,j} <t_i^z t_j^z>
\ \simeq \ {b \over 3 k_B T} {1+\overline{f} \over 1-\overline{f}} \ .
\label{corr}
\ee
In principle, this quantity depends on the average of $f(\psi)$, and
is not directly related to $\xi_{eff}$ (which depends on $\overline{\log(f)}$).
This property of non self-averageness of the correlation function has
been well studied in the litterature on disordered spin chains\cite{derrida}.
However it turns out that in the continuum limit $b\to 0$ this subtlety
disappears: using (\ref{fexpansion}), we find that the slope is related 
to the effective persistence length through the usual WLC formula 
$S=2 \xi_{eff}/(3 k_B T)$.

The analytical results just presented are limited to the low force
(linear response) regime.  To study the elastic behaviour of a
stretched heteropolymer for large extensions  ($ x \sim 1 $), we
will compute the partition function and the average extension
(magnetisation) by transfer matrix and Monte-Carlo (MC) methods.

The transfer matrix $T(\vec{t}_{i},\vec{t}_{i-1})$ can be read from
Eq.(\ref{e3}):

\begin{eqnarray}   \label{e9}
 T(\vec{t}_{i},\vec{t}_{i-1}) = \exp [ {K \cos (\alpha_i - \psi_i) +
H(\cos \theta_{i} + \cos \theta_{i-1})/2}] \ ,
\end{eqnarray}
with:
\begin{eqnarray} \label{e10}
\cos(\alpha_i - \psi_i) &=& \cos \psi_i ~ \vec{t}_{i} \cdot \vec{t}_{i-1}
+ \sin \psi_i ~ \sqrt{ 1 - (\vec{t}_{i} \cdot \vec{t}_{i-1})^2} \nonumber \\
\vec{t}_{i} \cdot \vec{t}_{i-1}&=& 
\cos \theta_{i} \cos \theta_{i-1}  + \sin \theta_{i} \sin \theta_{i-1} 
\cos ( \phi_{i} - \phi_{i-1}) \ .
\end{eqnarray}

The probability density  $\rho_i$ of the angles $\theta_i,\phi_i$ obeys
the recursion relation:
\begin{eqnarray}  \label{e11}
\rho_i(\theta_{i},\phi_{i}; \theta_0, \phi_0) = C  \int \sin \theta_{i-1}
d \theta_{i-1} d \phi_{i-1} ~ T(\vec{t}_{i},\vec{t}_{i-1}) ~ 
\rho_{i-1}(\theta_{i-1},\phi_{i-1}; \theta_0, \phi_0) \ ,
\end{eqnarray}
where $C$ is a normalisation constant.
The relative extension (magnetisation) of long chains is given
by $x = \lim_{N \to \infty}{\rm Tr}[\cos \theta \ \rho_N]$. 
 By discretising the angular variables $\{ \theta_i, \phi_i
\}$ and iterating Eq.(\ref{e11}), it is then straightforward to
calculate numerically the relative extension $x$ of a random chain due
to a force $F$. In the limit of long chains this characteristics
becomes independent of the particular realization of the disorder.
 The results are presented in Figs. 2 and 3, for the case
of half gaussian disorder with two different values of $r$.
The value of $b/\xi_T=.05$ has been chosen small enough that the 
results are close to those of the continuum limit (in the absence of disorder
the relative difference is less than 8 \%). The
continuous line is the force versus extension curve of a {\it homogenous}
WLC, Eq.(\ref{e2a}) with persistence length $\xi_{eff}$ given by
Eq.(\ref{xieff}). The remarkable and surprising agreement between the
numerical data and the WLC results implies that for all experimental
purposes the elongation curve of a random Kratky-Porod chain is identical
to that of a homogenous chain
with a smaller persistence length, i.e. it
follows Eq.(\ref{e2a}) with $\xi_{eff}$ replacing $\xi_T$.

\begin{figure}
\centerline{
\epsfxsize=8cm
\epsffile{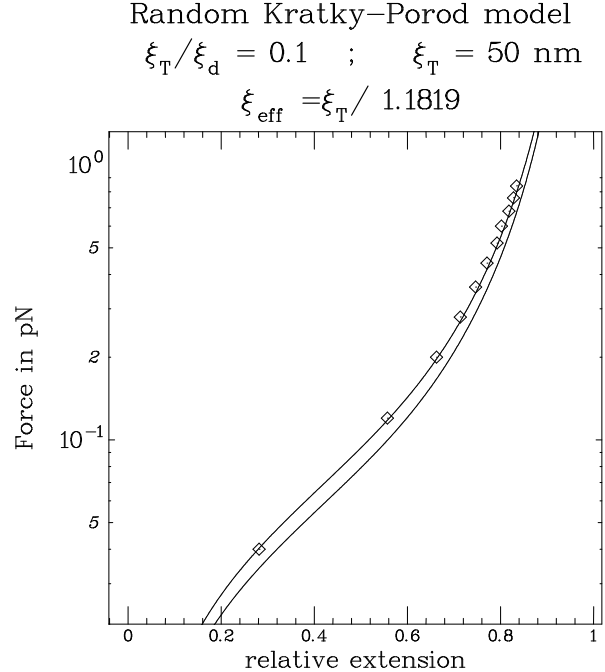}}
\caption{The force elongation characteristics of the random chain,
obtained by transfer matrix computations. The full lines are the predictions
of the model without disorder, with its original elastic constant
 (lower curve), and  with an effective elastic constant given by (\ref{xieff}) (upper
curve).}
\label{fig2}
\end{figure}

To further check this surprising result, we have run a Monte-Carlo
simulation of a discretised random Kratky-Porod chain with $b/\xi_T=.05$; $N=600$ ;
 ($k_B T = 1$). Rapid equilibration is
achieved by a MC move which involves a possible global rotation of the
chain for all segments $j \ge i$ with $i$ chosen at random). If the
energy of the chain after the move is lower than before it
(${\cal{E}}_{new} < {\cal{E}}_{old}$) the move is accepted with
probability $p=1$, otherwise it is performed with probability $p =
\exp[ -({\cal{E}}_{new} - {\cal{E}}_{old})]$. The total number
of Monte Carlo steps was around $2. \ 10^6$.In the absence of
disorder ($1/\xi_d = 0$) onr recovers the results of an homogenous WLC,
Eq.(\ref{e2a}). In the presence of disorder ($\xi_T/\xi_d = 2$) and after
averaging over $M_d = 80$ configurations of disorder, one recovers
the transfer matrix results for the extension of a random Kratky-Porod
chain (see Fig.3).

In summary, we have shown that a random Kratky-Porod chain with a certain type
of disorder is well aproximated by a pure chain with an effective elastic constant.
It would be interesting to check further the precision of this result, and
its dependance on $b$, to see if it holds for wider classes of disorder,
and to understand better why there is such a form of universality.

 We wish to thank R. Zeitak for useful discussions.
D.B. acknowledges the partial support of a Michael
fellowship.
D.D. acknowledges partial support from the
International Center for Theoretical Physics (Trieste) and the
laboratoire de Physique Statistique de l' ENS.

\begin{figure}
\centerline{
\epsfxsize=8cm
\epsffile{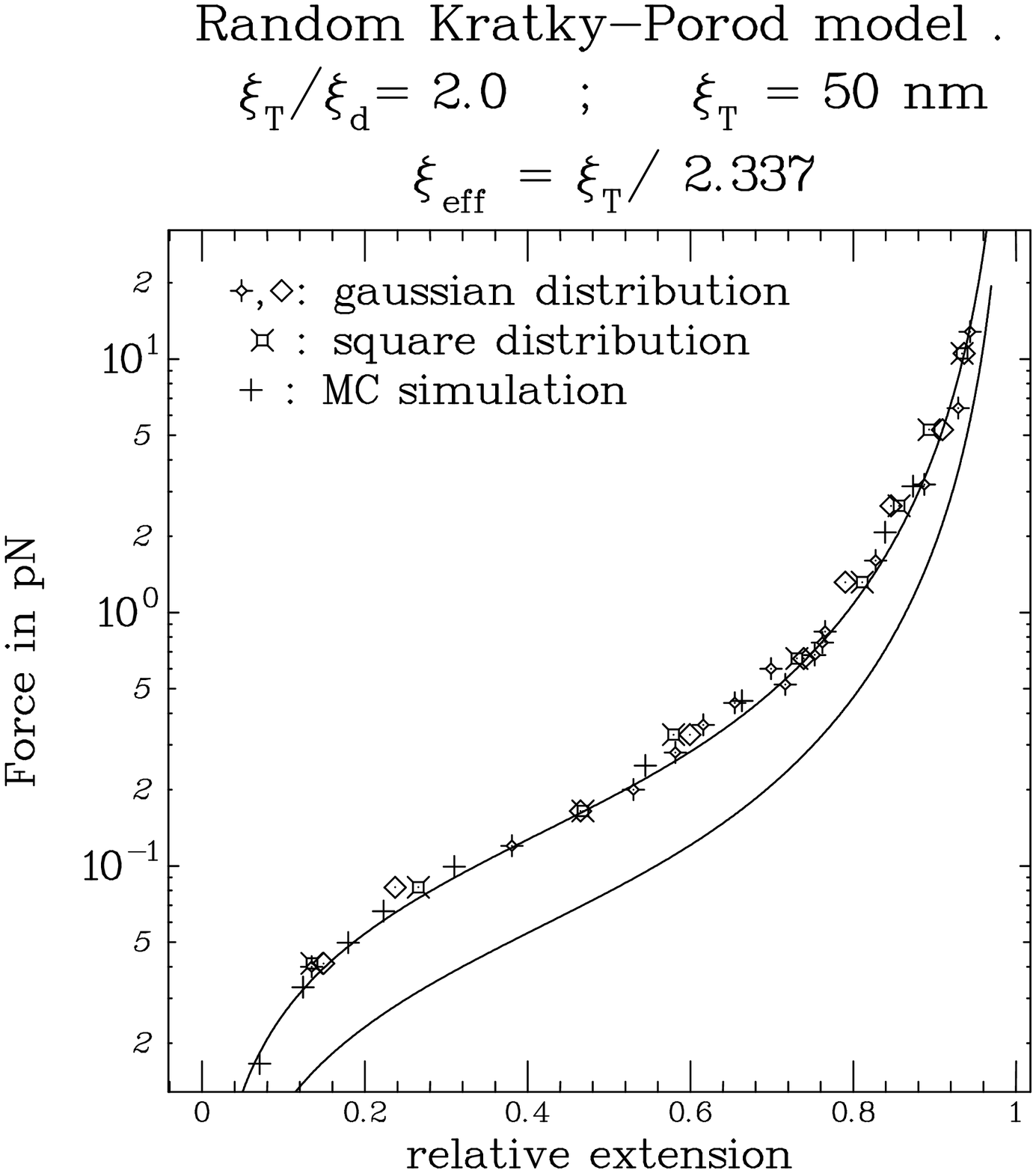}}
\caption{The force elongation characteristics of the random chain,
obtained by transfer matrix computations, and by Monte Carlo simulations.
 The full lines are the predictions
of the model without disorder, with its original elastic constant
 (lower curve), and  with an effective elastic constant $\xi_{eff}$
 given by (\ref{xieff}) (upper
curve). Two types of disorder, with the same $\xi_{eff}$, have been used in the
transfer matrix computation: a half gaussian distribution
of the angle $\psi$ (see text) and a flat one.}
\label{fig3}
\end{figure}

\end{document}